\documentclass[prc,twocolumn,epsfig,nofootinbib,floatfix]{revtex4}
\usepackage{graphics}
\usepackage{epsfig}
\usepackage{amsfonts}
\usepackage{amsmath}
\usepackage{bm}

\usepackage{color}

\begin{document}
\title{Toroidal, compression, and vortical dipole strengths in $^{124}$Sn \\ }
\author{J. Kvasil$^{1}$,  V.O. Nesterenko  $^{2}$, A. Repko  $^{1}$, W. Kleinig  $^{2,3}$,
P.-G. Reinhard  $^{4}$, and  N. Lo Iudice $^{5,6}$}
\affiliation{$^{1}$ \it  Institute of Particle and Nuclear Physics, Charles University,
CZ-18000, Prague, Czech Republic}
{\email{kvasil@ipnp.troja.mff.cuni.cz}
\affiliation{$^{2}$   \it Laboratory of Theoretical Physics, Joint Institute for Nuclear Research,
Dubna, Moscow region, 141980, Russia}
\email{nester@theor.jinr.ru}
\affiliation{$^{3}$  \it Technische Universit\"at Dresden,
  Institut f\"ur Analysis, D-01062, Dresden, Germany}
\affiliation{$^{4}$
  \it Institut f\"ur Theoretische Physik II,
     Universit\"at Erlangen, D-91058, Erlangen, Germany}
\affiliation{$^5$ Dipartimento di Scienze Fisiche,
Universit$\grave{a}$ di Napoli Federico II, Monte S Angelo, Via
Cinzia I-80126 Napoli, Italy}
\affiliation{ $^6$ Istituto
Nazionale di Fisica Nucleare, Monte S Angelo, Via Cinzia I-80126
Napoli, Italy}

\begin{abstract}
  The toroidal, compression and vortical dipole strength functions
in semi-magic $^{124}$Sn (and partly in doubly-magic
$^{100,132}$Sn) are analyzed  within the
random-phase-approximation method with the SkT6, SkI3, SLy6,
SV-bas, and SkM$^*$ Skyrme forces. The isoscalar (T=0), isovector
(T=1), and electromagnetic ('elm') channels are considered. Both
convection $j_c$ and magnetization $j_m$ nuclear currents are
taken into account. The calculations basically confirm the
previous results obtained for $^{208}$Pb with the force SLy6. In
particular, it is shown that the vortical and toroidal strengths
are dominated by $j_c$ in T=0 channel and by $j_m$  in T=1 and
'elm' channels. The compression strength is always determined by
$j_c$. Is also shown that the 'elm' strength (relevant for (e,e')
reaction) is very similar to T=1 one. The toroidal mode resides in
the region of the pygmy resonance. So, perhaps, this region
embraces both irrotational (pygmy) and vortical (toroidal) flows.
\end{abstract}

\maketitle

\section{Introduction}
The toroidal and compression flows represent  a subject of intense
investigation for many years, see e.g. the review \cite{Pa07}. The
toroidal mode (TM) is known to be determined by the second order
correction to the leading long-wave part of the transition
$E\lambda$ operator \cite{Du75,Kv03,Kv11}. The compression mode
(CM) \cite{Ha77,Se81,St82}, being different from TM by
construction, is related, nevertheless, to TM and also represents
the second-order correction \cite{Kv11,Se81}. Despite the long
studies, the vorticity of the modes is still a subject of
discussions \cite{Kv11}. Besides, a possible coexistence of the
dipole TM and pygmy mode (PM) becomes actual \cite{Ry02}.

In hydrodynamics (HD),  the vorticity is defined by the curl of
the velocity field \cite{La87}. Then TM is vortical and CM is
irrotational \cite{Kv11}. However, nuclear models deal with the
nuclear current rather than the velocity field, which causes
alternative definitions of the vorticity. In \cite{Ra87}, the
nuclear current component $j^{(fi)}_{\lambda \: l=\lambda+1}(r)$,
arising in the multipole decomposition of the current transition
density $<f| \hat{\vec{j}}_{\mathrm{nuc}}(\vec{r}) |i>$,  is
treated as unconstrained by the continuity equation and thus
suitable as a measure of the vorticity. Following this definition,
the TM and CM are mixed flows with both vortical and irrotational
elements. In our recent study \cite{Kv11}, the relevant vortical
operator was derived and related by a simple manner to the CM and
TM operators. The vortical, toroidal and compression E1 strengths
were numerically explored with the separable
random-phase-approximation (SRPA) \cite{Ne02,Ne06} in $^{208}$Pb
with the force SLy6 \cite{Ch97}.

In this paper, the toroidal, compression and vortical E1 strengths
are investigated in semi-magic $^{124}$Sn within a wide range of
Skyrme forces (SkT6 \cite{To84}, SV-bas \cite{Kl09},
SkM*\cite{Ba82}, SLy6 \cite{Ch97}, SkI3 \cite{Re95}). In addition
to T=0 and T=1 channels, the electromagnetic ('elm') strength
relevant for (e,e') reaction is inspected. The trends with  the
neutron number, from $^{100}$Sn to $^{132}$Sn, are considered. A
possible relation of TM, CM, and PM is discussed.

\section{Model and calculation scheme}
The connection of the vortical, toroidal, and compression
transition operators with the standard electric multipole operator $\hat{M}(E\lambda\mu, k)$
is discussed in detail in \cite{Kv11}.
The toroidal operator $\hat{M}(\mathrm{tor}, \lambda\mu)$ appears \cite{Du75,Kv03,Kv11}
as a second order term in the decomposition of $\hat{M}(E\lambda\mu, k)$ in terms
of momentum transfer $k$:
\begin{equation}\label{2}
\hat{M}(E\lambda \mu;\:k) = \hat{M}(E \lambda \mu)
+ k\;\hat{M}(\mathrm{tor}, \lambda \mu) +\ldots
\end{equation}
where
\begin{eqnarray} \label{3}
&& \hat{M}(E\lambda \mu) = - \int \: d^3r \:\hat{\rho}(\vec{r})
\:r^{\lambda} \: Y_{\lambda \mu}(\hat{\vec r}) \; ,
\\
&& \hat{M}(\mathrm{tor}, \lambda \mu) = -
\frac{i}{2c}\:\sqrt{\frac{\lambda}{2\lambda+1}} \int \: d^3r
\hat{\vec{j}}_{\mathrm{nuc}}(\vec{r}) \: r^{\lambda+1}
\nonumber \\
&& \quad \cdot \left[ \:\vec{Y}_{\lambda \:\lambda-1\:\mu}(\hat{\vec r})
 + \sqrt{\frac{\lambda}{\lambda+1}}
\:\frac{2}{2\lambda+3} \:\vec{Y}_{\lambda \:\lambda+1\:\mu}(\hat{\vec r}) \:\right]
\end{eqnarray}
are long-wave limits of the standard electric and toroidal
operators, respectively; $\vec{Y}_{\lambda l \mu}(\hat{\vec r})$
is the vector spherical harmonic, and $\hat{\rho}(\vec{r})$ is the
density operator.

Following the concept \cite{Ra87}, the vortical operator is
\cite{Kv11}
\begin{eqnarray} \label{4}
&& \hat{M}(\mathrm{vor}, \lambda \mu) =
 - \frac{i}{c} \:\frac{1}{2\lambda+3}\:\sqrt{\frac{2\lambda+1}{\lambda+1}}\:
\int \: d^3r \:  \hat{\vec{j}}_{\mathrm{nuc}}(\vec{r})
\nonumber \\
&& \qquad \qquad \qquad \qquad \qquad \:\cdot \: r^{\lambda+1}
\:\vec{Y}_{\lambda \: \lambda+1 \:\mu}(\hat{\vec r})
\end{eqnarray}
and its transition matrix elements serve as a measure of the
vorticity for a given excitation.

The vortical, toroidal, and compression operators  are simply
coupled  \cite{Kv11}:
\begin{equation} \label{5}
\hat{M}(\mathrm{vor}, \lambda \mu) = \hat{M}(\mathrm{tor}, \lambda \mu)
 + \hat{M}(\mathrm{com}, \lambda \mu) \; ,
\end{equation}
where
\begin{eqnarray} \label{6}
&& \hat{M}(\mathrm{com}, \lambda \mu) = - k \: \hat{M}(\mathrm{com'}, \lambda \mu)
\nonumber \\
&&
 = - \frac{i}{2c} \: \frac{1}{2\lambda+3} \: \int \:d^3r \: r^{\lambda+2} \:
Y_{\lambda \mu}(\hat{\vec r}) \: \vec{\nabla} \cdot
\hat{\vec{j}}_{\mathrm{nuc}}(\vec{r})
\end{eqnarray}
is the current-dependent compression operator related to its
familiar density-dependent counterpart \cite{Se81}
\begin{equation} \label{7}
\hat{M}(\mathrm{com'}, \lambda \mu) =
\frac{1}{2\:(2 \lambda +3)} \:\int \:d^3r \:\hat{\rho}(\vec{r}) \:
r^{\lambda+2} \: Y_{\lambda \mu}(\hat{\vec r}) \; .
\end{equation}
In (\ref{3})-(\ref{6}), the symbol
$\hat{\vec{j}}_{\mathrm{nuc}}(\vec{r})$ stands for the operator of
the nuclear current embracing the convectional and magnetization
parts:
\begin{equation}\label{8}
\hat{\vec{j}}_{\mathrm{nuc}}(\vec{r}) = \hat{\vec{j}}_{c}(\vec{r})
+ \hat{\vec{j}}_{m}(\vec{r}).
\end{equation}
Explicit expressions for $\hat{\vec{j}}_{c}(\vec{r})$ and
$\hat{\vec{j}}_{m}(\vec{r})$ can be found elsewhere, see e.g.
\cite{Ri80}.

In this paper, only dipole transitions g.s. $\to I^{\pi}=1^-$ are considered.
Taking into account
the center-of-mass corrections (c.m.c.) in T=0 channel, the dipole operators
to be used in the calculations read \cite{Kv11}
\begin{equation} \label{11}
 \hat{M}(\mathrm{vor}, 1\mu) =
 -\frac{i}{5c}\:\sqrt{\frac{3}{2}}\:\int\:d^3r \: \hat{\vec{j}}_{\mathrm{nuc}}(\vec{r}) \cdot
r^2\:\vec{Y}_{12\mu}(\hat{\vec r}) \;,
\end{equation}
\begin{eqnarray} \label{9}
&& \hat{M}(\mathrm{tor}, 1\mu) = -
\frac{2}{2c\sqrt{3}}\:\int\:d^3r \:
\hat{\vec{j}}_{\mathrm{nuc}}(\vec{r})
\nonumber \\
&&  \cdot\left[\:\frac{\sqrt{2}}{5} \:r^2 \:\vec{Y}_{12\mu}(\hat{\vec r}) +
(r^2 - \delta_{T,0} \langle r^2\rangle_0)\:\vec{Y}_{10\mu}(\hat{r})\:\right] ,
\\
\label{10}
&&\hat{M}(\mathrm{com'}, 1\mu) = \frac{1}{10}\:\int\:d^3r \: \hat{\rho}(\vec{r})
\nonumber \\
&& \qquad \quad \cdot\:
\left[\: r^3 - \delta_{T,0} \:\frac{5}{3}\:\langle r^2\rangle_0 \:r\:\right]\:
Y_{1\mu}(\hat{\vec r})
\end{eqnarray}
where $\langle r^2\rangle_0 = \int\:d^3r \: \rho_0(\vec{r})$ is the ground state squared radius.
Note that the vortical operator has no the c.m.c..

The calculations were performed within the  SRPA \cite{Ne02,Ne06}.
The model is fully self-consistent since both the mean field and
residual interaction are derived from the same Skyrme functional.
Moreover, the residual interaction includes all the functional
contributions and the Coulomb (direct and exact) terms. The
self-consistent factorization of the residual interaction
dramatically reduces the computational effort while keeping the
accuracy of nonseparable RPA. SRPA has been successfully applied
to description of electric
\cite{Kv11,nest_IJMPE_07_08,nest_PRC_08,Kva_IJMPE_12} and magnetic
\cite{Ve09,Nest_JPG_10,Nest_IJMPE_10} giant resonances as well as
E1 strength near the particle thresholds
\cite{Kva_IJMPE_09,Kva_IJMPE_11}.
\begin{figure}
\includegraphics[width=7.0cm]{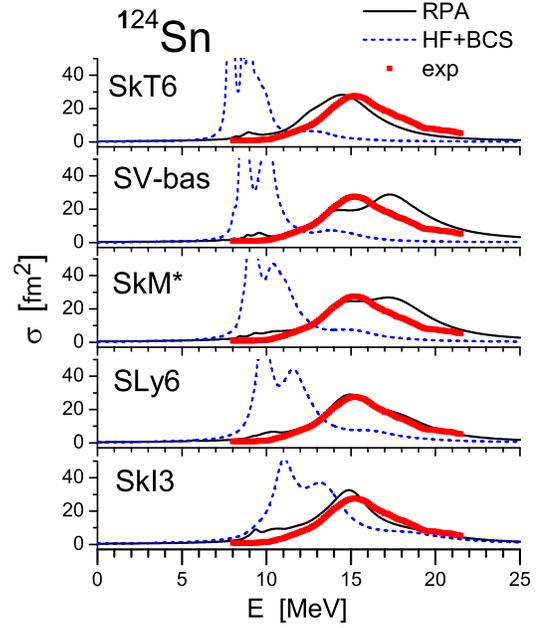}
\caption{\label{fig1} (color online) Photoabsorption cross section
in $^{124}$Sn for different Skyrme parametrizations. The SRPA and
HF+BCS cross sections are compared to the experimental data
\protect\cite{Va03}.}
\end{figure}

For all the modes, the strength function reads
\begin{equation}\label{SF}
  S_{\alpha}(E1; E) = 3
\sum_{\nu}
  |\langle\Psi_\nu|\hat{M}_{\alpha}(E10)|\Psi_0\rangle|^2
  \zeta(E - E_{\nu})
\end{equation}
where $\zeta(E - E_{\nu})$ is a Lorentzian weight with the averaging parameter
$\Delta$=1 MeV; $\hat{M}_{\alpha}(E1\mu)$ is the transition operator of the type
$\alpha =\{\mathrm{vor, tor, com, com'}\}$; $\Psi_0$ is the ground state wave function;
$E_{\nu}$ and $|\Psi_\nu\rangle$ are the energy and wave function of
the RPA $\nu$-state.

The T=0, T=1, and 'elm' channels are defined by the proper choice
of the proton and neutron effective charges
$e_{n,p}^{\mathrm{eff}}$ and gyromagnetic factors
$g_{n,p}^{\mathrm{eff}}$ as:
\begin{eqnarray} \label{15}
T=0: && e_n^{\mathrm{eff}}=e_p^{\mathrm{eff}}=1,
\quad g_{n,p}^{\mathrm{eff}}=\frac{\zeta}{2}\:(g_n + g_p) \;,
\\
\label{16}
T=1: && e_n^{\mathrm{eff}}=-e_p^{\mathrm{eff}}=-1, \;
g_{n,p}^{\mathrm{eff}}=\frac{\zeta}{2}\:(g_n - g_p) \;,
\\
\label{17}
el: && e_n^{\mathrm{eff}}=0,\: e_p^{\mathrm{eff}}=1,
\quad g_{n,p}^{\mathrm{eff}}=\zeta g_{n,p} \;.
\end{eqnarray}
Here $g_{n} = -3.82$ and $g_{p} = 5.58$ are free neutron and proton
gyromagnetic ratios; $\zeta \approx 0.7$  is the quenching factor.

For neutrons in the semi-magic nucleus $^{124}$Sn, the zero-range pairing forces
are used at the BCS level (HF+BCS) \cite{Ben00}. More details of the calculations
are found in \cite{Kv11}.


\section{Results and discussions}

\begin{figure} 
\includegraphics[width=7cm]{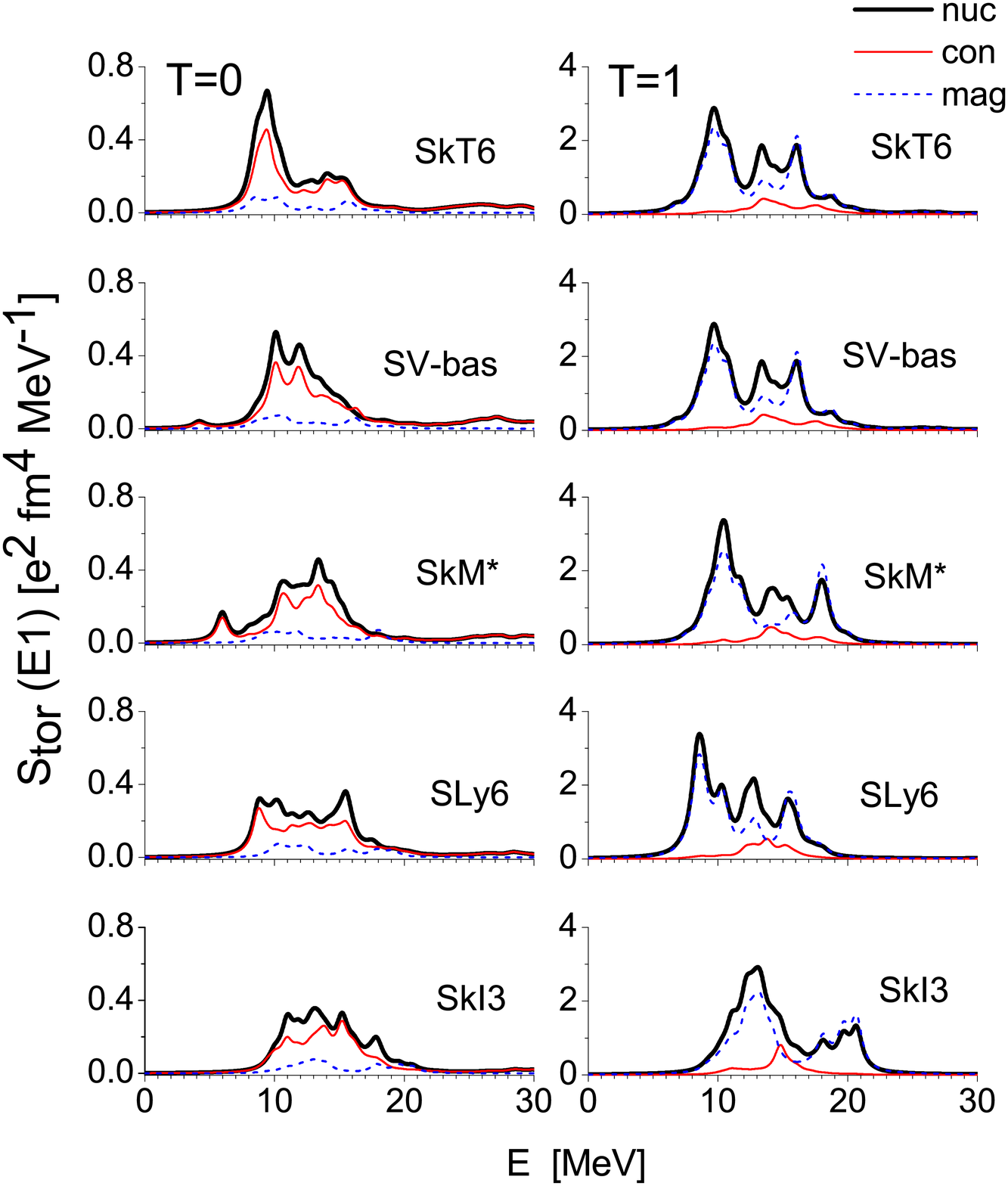}
\caption{\label{fig2}
 (color online) Toroidal strength function in
$^{124}$Sn for T=0 (left panel) and T=1 (right panel) channels,
computed with the total (black bold line) , convection (red thin
line), and magnetization (blue dash line) nuclear current.
Different Skyrme forces are used as indicated.}
\end{figure}
\begin{figure} 
\includegraphics[width=7cm]{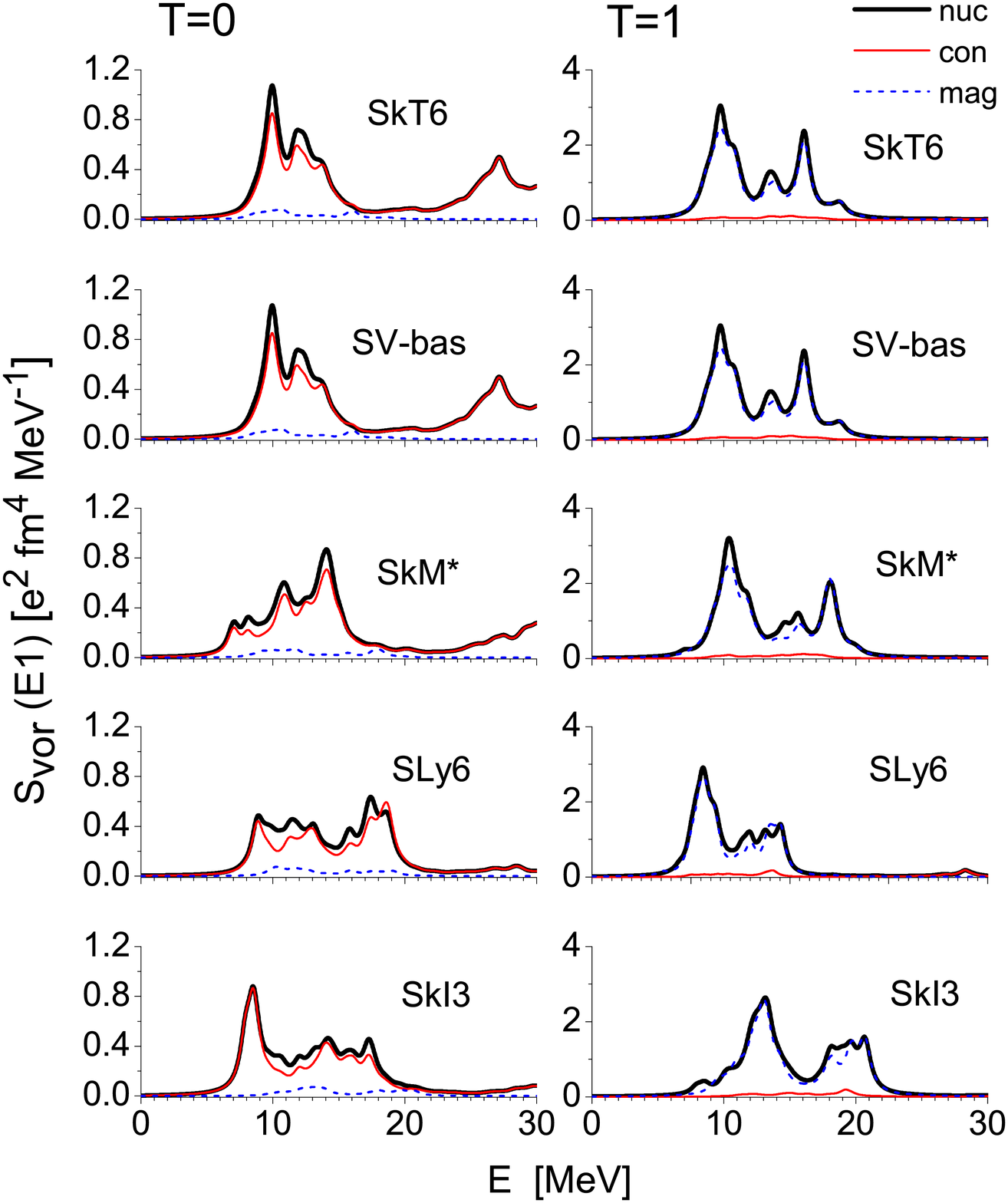}
\caption{\label{fig3}
(color online) The same as in Fig.2 but for
the vortical strength.}
\end{figure}
\begin{figure} [t] 
\includegraphics[width=6cm]{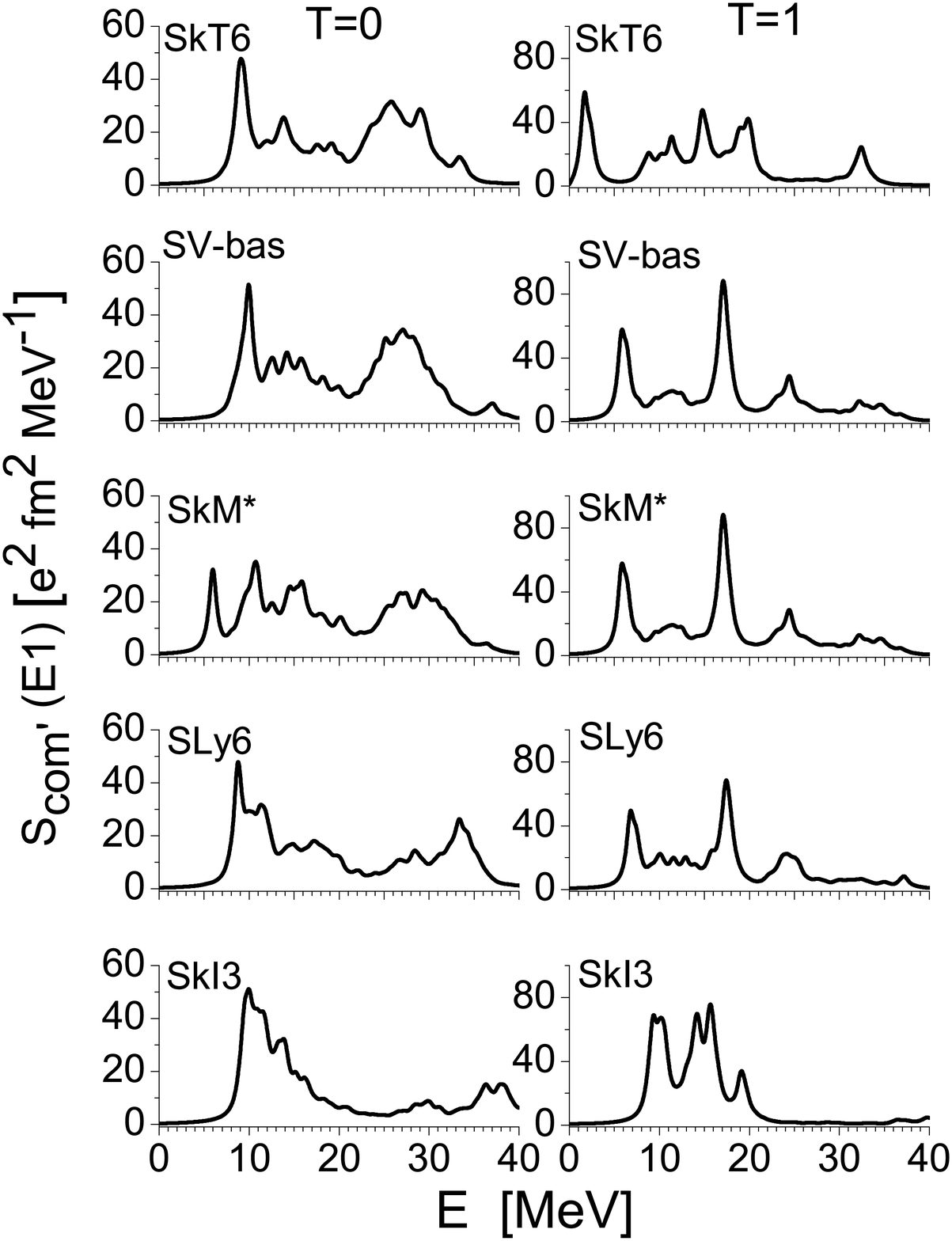}
\caption{\label{fig4}
(color online) The compression strength in
$^{124}$Sn for T=0 (left panel) and T=1 (right panel) channels,
generated by the operator (\ref{10}). Since $j_m$-contribution is
zero, only $j_{\mathrm{nuc}}=j_c$ case is plotted.}
\end{figure}
\begin{figure} 
\includegraphics[width=7cm]{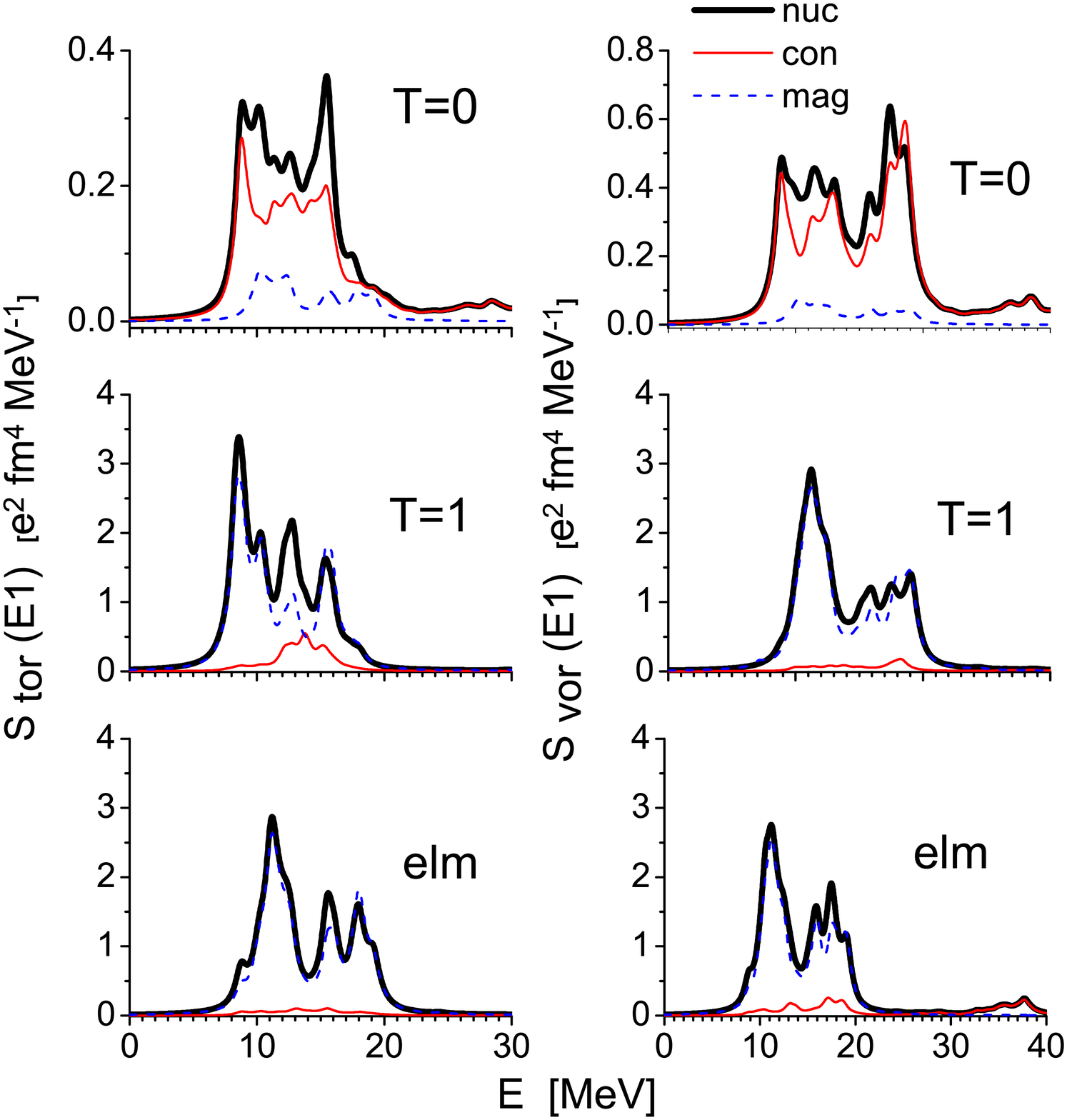}
\caption{\label{fig5} (color online) Toroidal (left) and vortical
(right) strengths in $^{124}$Sn, computed  with the total (black
bold line), convection (red thin line), and magnetization (blue
dash line) nuclear current. The T=0 (upper plot), T=1 (middle
plot), and electromagnetic (bottom plot) channels are considered.
Only the force SLy6 is used.}
\end{figure}
\begin{figure} [t] 
\includegraphics[width=6.0cm]{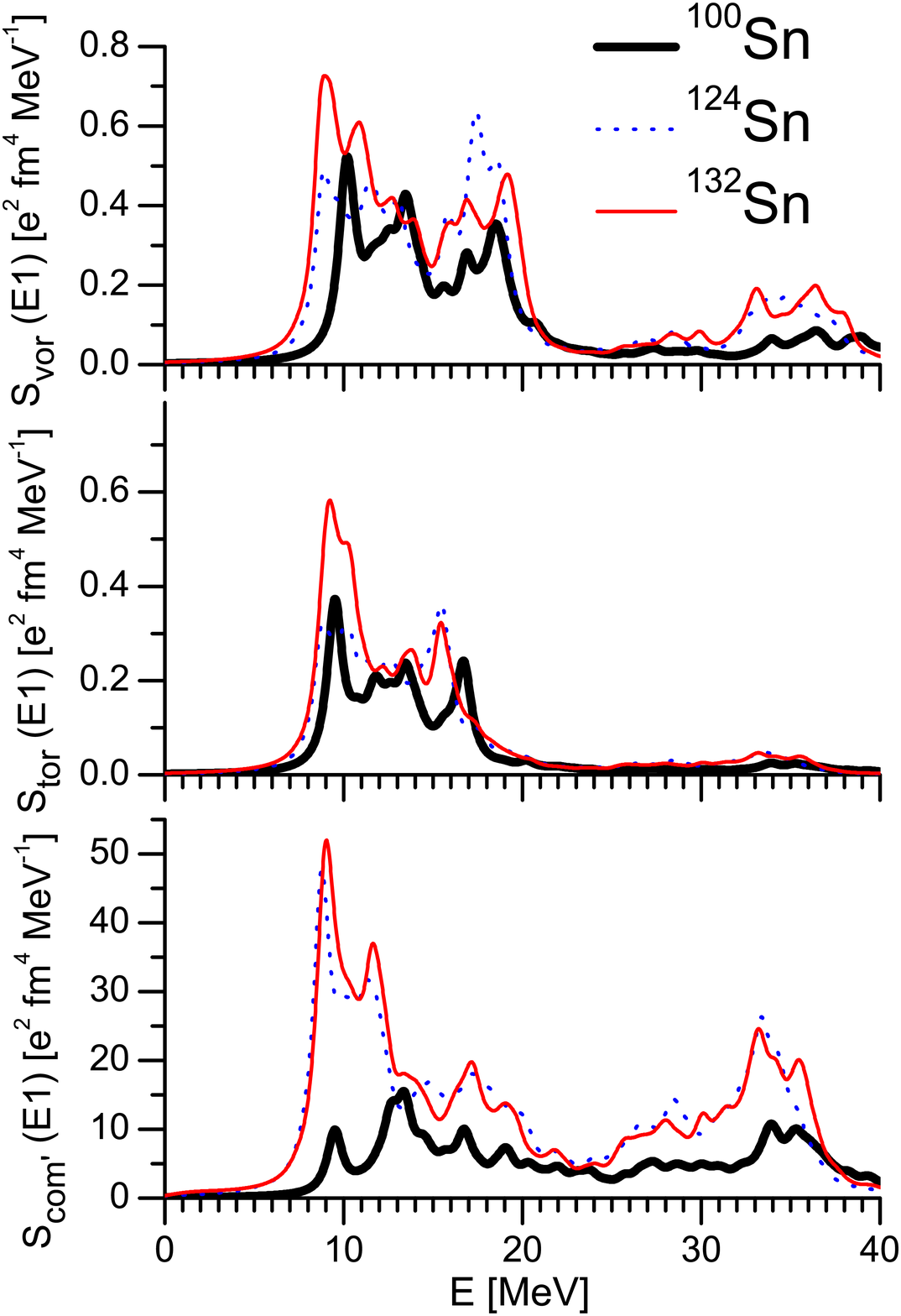}
\vspace{3mm}
 \caption{\label{fig7} The vortical, toroidal, and
compression strength functions in T=0 channel in
$^{100,124,132}$Sn isotopes, calculated within SRPA with the force
SLy6.}
\end{figure}

Results of the calculations are given in Figs. 1-6. In Fig. 1, the
SRPA accuracy is inspected for the photoabsorption in $^{124}$Sn.
Unlike the strength functions (\ref{SF}), the photoabsorption
$\sigma$ is computed with an energy-dependent averaging parameter
$\Delta (E)$ in the Lorentzian weight, for more details see
\cite{Kva_IJMPE_11}. A representative set of Skyrme forces with
various isoscalar effective masses is used: SkT6 ($m^*$=1), SV-bas
($m^*$=0.9), SkM$^*$ ($m^*$=0.79), SLy6 ($m^*$=0.69), and SkI3
($m^*$=0.58). Both SRPA and HF+BCS results are depicted.

Figure 1 shows a good agreement with the experimental data
\cite{Va03} for all the forces.
The best agreement is for SLy6. A discrepancy with
the experiment around the particle emission threshold (9-10 MeV)
may be explained by neglecting the coupling with complex
configurations, which is expected to be strong in this particular
energy region \cite{Li08,En10}. Note a small energy upshift of the
double-peak HB+BCS strength with decreasing the effective mass
$m^*$ from SkT6 to SkI3 (this may be explained by spreading the
mean-field spectra with $m^*$).  As seen from Fig. 1, the residual
interaction drastically shifts the strength to a higher energy,
neutralizes the $m^*$ effect, and transforms the double-peak
structure to a single-peak one.

Figures 2 and 3 show the toroidal and vortical strength functions in $^{124}$Sn for T=0 and
1 channels and the same set of Skyrme forces. The cases of the total (\ref{8}), convection, and
magnetization nuclear currents are considered. It is seen that, in agreement with the previous
SLy6 results  for $^{208}$Pb \cite{Kv11}, the isoscalar (isovector) strengths are dominated by
the convection $j_c$ (magnetization $j_m$) current. This is explained by the destructive
(constructive) interference of the neutron and proton $j_m$-contributions  in T=0 (T=1)
channels.

As mentioned in Sec. 2, the toroidal and vortial strengths
represent the second-order E1 corrections to the leading
first-order E1 response (e.g. photoabsorption).  These
second-order strengths are less collective \cite{Kv11} and,
following Figs. 2-3, more sensitive to the Skyrme
parameterization. It is seen that, though all the Skyrme forces
give qualitatively similar pictures, the quantitative description
of the toroidal and vortical strength functions (energy centroids,
widths, and gross-structure) noticeably depends on the force. In
particular, the SRPA energy centroids are generally upshifted with
$m^*$, like in the HF+BCS photoabsorption in Fig. 1. Note that
dependence on the isoscalar effective mass $m^*$ takes place in
both T=0 and 1 channels.

It is remarkable that the region 6-10 MeV,  often related to the
pygmy resonance, embraces noticeable fractions of the toroidal and
vortical strengths, especially in T=0 channel. For some forces,
e.g. SLy6, the strengths are even peaked there. This hints that
the region 6-10 MeV may host both the vortical (toroidal) and
irrotational (pygmy) motions.

In Fig.4, the compression strength function in $^{124}$Sn, generated by
the operator (\ref{10}), is presented. The CM has no contribution from the
magnetization current $j_m$ and is fully determined by the convection current $j_c$ \cite{Kv11}.
It is seen that, like in the previous figures for the toroidal and vortical strengths,
the CM demonstrates for T=0,1 a significant dependence on the Skyrme
parameterization. The upshift  of the strength with $m^*$ (though not regular) is also
visible.

In Fig. 5, the compositions of the toroidal and vortical strengths
in T=0 and 1 channels (\ref{15})-(\ref{16}) are compared  to the
'elm' channel (\ref{17}) relevant for (e,e') reaction. It seen
that T=1 and 'elm'  responses are very similar and both mainly
determined by the magnetization current. Thus we encounter a
remarkable case when the electric modes are driven not by $j_c$
but $j_m$, and this case may in principle be observed in (e,e')
reaction.

The isotopic dependence of isoscalar toroidal, vortical, and
compression strengths is illustrated in Fig. 6. In addition to
$^{124}$Sn,  the neutron-deficit $^{100}$Sn and neutron-excess
$^{132}$Sn doubly-magic nuclei  are considered. The  neutron
excess (skin) over the N=Z nuclear core is zero in $^{100}$Sn and
essential in $^{132}$Sn. Fig. 6 shows that the most strong
isotopic effect takes place for CM. There is a considerable growth
of the CM high-energy branch from $^{100}$Sn to $^{132}$Sn, which
may be related to increasing the neutrons participating in the
dipole compression. The CM low-energy branch located at the pygmy
resonance region 6-10 MeV grows from $^{100}$Sn to $^{132}$Sn even
more.
This indicates a strong dominance of neutron oscillations in the
region. Note that the dipole compression and familiar pygmy-like
(oscillation of the neutron skin against the N=Z core) flows are
both irrotational in the HD sense and their velocity fields look
rather similar. So their mixture in the pygmy resonance region
seems natural. Figure 6 also shows that, unlike the CM, the
toroidal and vortical strengths weakly depend on the neutron skin.
Perhaps the vortical (toroidal) motion takes place mainly in
nuclear N=Z core. The remaining minor isotopic effect may be
related to the coupling between the neutron oscillations and
toroidal internal motion. Anyway, both these collective motions
share about the same energy region.


\section{Conclusion}
The toroidal, vortical, and compression dipole strength functions
in $^{100,124,132}$Sn isotopes were analyzed in the framework of
the self-consistent separable Skyrme-RPA approach
\cite{Ne02,Ne06}. A representative set of five Skyrme forces with
essentially different isoscalar effective mass $m^*$ was used. The
isoscalar (T=0), isovector (T=1), and electromagnetic ('elm')
channels were inspected. The calculations generally confirm the
previous results for $^{208}$Pb obtained for the Skyrme force
SLy6. In particular, it was corroborated that the toroidal and
vortical strengths in T=0 (T=1) channels are mainly provided by
the convection (magnetization) nuclear current. A close similarity
of T=1 and 'elm' channels was established. This means the E1(T=1)
toroidal mode represents an unusual case of the electric
collective motion determined by the magnetization current and this
case can in principle be explored in (e,e') reaction.

The comparison of the results for $^{100,124,132}$Sn shows that
both low- and high-energy branches of the compression mode
considerably depend on the neutron excess. The enhanced dipole
strength in the low-energy region, often called pygmy resonance,
is strongly correlated with the irrotational low-energy
compression mode and vortical toroidal mode (probably located in
the nuclear N=Z core).  Altogether, the pygmy resonance region
shows an impressive coexistence of various irrotational and
vortical flows.

\section*{Acknowledgments}
The work was partly supported by the GSI-F+E, Heisenberg-Landau
(Germany - BLTP JINR), and Votruba - Blokhintsev (Czech Republic - BLTP JINR)
grants. W.K. and P.-G.R. are grateful for the BMBF support under contracts 06
DD 9052D and 06 ER 9063. The support of the research plan MSM 0021620859
(Ministry of Education of the Czech Republic) is also appreciated.


\end{document}